\documentclass[prd,floatfix,showpacs,twocolumn,amsmath,amssymb]{revtex4}
\usepackage{natbib}
\usepackage{dcolumn}
\usepackage{graphicx}
\usepackage{color}
\usepackage{bm}
\newcommand{\be}{\begin{equation}}
\newcommand{\ee}{\end{equation}}
\newcommand{\bea}{\begin{eqnarray}}
\newcommand{\eea}{\end{eqnarray}}

\def\[{\begin{equation}}
\def\]{\end{equation}}

\begin{document}
\title{Cross-correlation between cosmic microwave background lensing and galaxy intrinsic alignment as a contaminant to gravitational lensing cross-correlated probes of the universe}
\author{M. A. Troxel\footnote{Electronic address: troxel@utdallas.edu}}
\author{Mustapha Ishak\footnote{Electronic address: mishak@utdallas.edu}}
\affiliation{
Department of Physics, The University of Texas at Dallas, Richardson, TX 75080, USA}
\date{\today}
\begin{abstract}
We introduce here a cross-correlation term between CMB lensing and galaxy intrinsic alignment, noted hereby as $\phi I$. This effect acts as a contaminant to the cross-correlation between CMB lensing and galaxy lensing. The latter cross-correlation has recently been detected for the first time, and measurements will greatly improve as the area of overlap between galaxy and CMB surveys increases and measurements of the CMB polarization become more significant. This will constitute a powerful probe for studying the structure and evolution of the universe.
The magnitude of the $\phi I$ term is found to be about 15\% of the pure CMB lensing-galaxy lensing component and acts to reduce the magnitude of its measured spectrum. This offset in the spectrum will strongly impact its use for precision cosmological study if left unmitigated. We also propose here a method to calibrate this $\phi I$ contamination through use of a scaling relation that allows one to reduce the impact of $\phi I$ by a factor of 20 or more in all redshift bins, which would reduce its magnitude down to detection limits in almost all cases. This will allow the full use of this probe for precision cosmology.   
\end{abstract} 
\maketitle
\section{Introduction}
{Gravitational lensing due to intervening large scale structure of the universe (cosmic shear) is a powerful probe for studying the structure and evolution the universe. It constrains, for example, the properties of dark energy and the amplitude of matter density fluctuations (e.g. \cite{1a,1b,1c,1d,1e,1f,1g,1h,1i,1j,1k,1l,Ishak,1m,1n,1o,1p}), and is a powerful test of gravity on large scales (\cite{2a,2b,2c,2d,2e,2f,2g,2h,2i,2j,2k,2l,2m,2n,2o,2p,2q} and references therein). Cosmic shear can be detected through the usual analysis of the weak lensing of galaxy shapes in large scale galaxy surveys, but also in surveys of the cosmic microwave background (CMB) (e.g. \cite{cmb1,cmb2,cmb3,cmb4,cmb5,cmb6,cmb7}). Such surveys are measuring for the first time the imprint of the gravitational lensing signature in the temperature fluctuations and polarization of the CMB, which place separate constraints from galaxy surveys on the matter fluctuation amplitude and the dark energy equation of state \citep{cmbl1,cmbl2,cmbl3,cmbl4,polarbear2013}. This allows us to utilize the CMB as a lensing source of known and fixed redshift to map the total intervening structure in the universe. These measurements will rapidly improve with better CMB polarization measurements, (e.g. see \cite{hu2002}), and can be used to calibrate systematics in shear measurements, (e.g. see \citep{Das2013b}).}

These probes are complementary, measuring the same information on large-scale gravitational potentials, and thus allow us to directly compare information about mass distributions in the universe across large redshifts through two independent methods, each with their own sources of error and systematics. Since these two probes measure the same lensing signal, they should be correlated. This cross-correlation has been detected by \cite{hand} with a significance of 3.2$\sigma$, which provides constraints on the amplitude of density fluctuations at intermediate redshifts (z=0.9), where the efficiency of the probes overlap. 

{Utilizing this cross-correlation in addition to the CMB lensing and galaxy lensing correlations provides a cleaner lensing signal, since the survey systematics of either CMB or galaxy survey should be uncorrelated \cite{Das2013b}. It also provides several physical benefits in constraining information about the evolution of structure in the universe. For example, the peak efficiency of the cross-correlation signal is significantly higher in redshift than that of a galaxy lensing survey, but intermediate to the peak lensing efficiency of the CMB. For the Canada-France-Hawaii Telescope Stripe 82 Survey (CS82) analyzed by \cite{hand} (see also references therein), this corresponds to a peak efficiency near the mean redshift $z=0.9$. This is about twice the peak efficiency for the CS82 galaxy lensing alone \cite{hand}, and probes instead structure at much larger redshift. Finally, it also introduces the ability to combine CMB and galaxy lensing information through tomographic cross-correlations across a wide range of redshifts. Thus combining CMB lensing, tomography in galaxy lensing, and the cross-correlation between the two opens up a wide range of redshifts within which to study information on structure in the universe.}

The most serious physical systematic in cosmic shear surveys is the intrinsic alignment of galaxies (IA) (e.g. \cite{7,8a,8b,8c,8d,8e,8f,8g,8h,9a,9b,10,hirata,12,13,14a,14b,15,16,17,6j,18b,19,blazek,heymans,merkel1,chisari,valageas}), where correlations involving the intrinsic shape of the galaxy before lensing contaminate the lensing signal and bias cosmological information. For the power spectrum, there exist two correlations involving the intrinsic alignment of the galaxies. First, there is a direct correlation ($II$) between the shapes of pairs of galaxies which reside or evolve within the same dark matter halo, and thus are tidally aligned causing a positive correlation in their shapes. The second correlation ($GI$), first identified by \cite{hirata}, is instead a correlation between the intrinsic shape of a foreground galaxy and the shape of a background galaxy that is lensed by the same structure which tidally aligns the first. This $GI$ correlation is an anti-correlation, and thus competes with the $II$ correlation in its impact on the lensing signal. While the $II$ correlation greatly decreases in magnitude as the physical separation of the two galaxies increases, and thus can be rendered negligible by utilizing cross-correlations between large redshift bins (see for example \cite{6k}), the $GI$ correlation can increase with separation and must be dealt with in more sophisticated ways. The $GI$ correlation has been measured in a variety of surveys (e.g. \cite{12,10,13,14a,14b,15}).

There exists no intrinsic alignment contamination in CMB lensing, which is one benefit to exploring its use in addition to galaxy lensing. However, the cross-correlation between CMB lensing and galaxy lensing will be contaminated by an intrinsic alignment correlation like $GI$, which we label $\phi I$. In the same way the intrinsic shape of a foreground galaxy can be correlated with the lensing of a background galaxy, a foreground galaxy can be correlated with the lensing deflection induced in the CMB temperature fluctuations or the polarization signal. This can be a potential source of bias, and we explore the expected magnitude of the $\phi I$ correlation as a fraction of the CMB lensing-galaxy lensing cross-correlation. We also extend the previously developed self-calibration methods (e.g. \cite{zhang,troxel,troxelc}) to propose a method for calibrating the $\phi I$ contamination by using information gained from galaxy weak lensing surveys. This work can be generalized to higher-order correlations, where there exist analogous contaminants such as $\phi II$, $\phi IG$, and $\phi\phi I$ in the bispectrum.

\section{Galaxy and CMB gravitational lensing formalism}
Under the Born approximation, the convergence $\kappa$ of a source galaxy at comoving distance $\chi_G$ and direction $\hat{\theta}$ in a flat, $\Lambda$CDM universe is related to the matter density contrast $\delta$ through the lensing kernel $W_L(\chi_L,\chi_G)=\frac{3}{2}\Omega_m(1+z_L)\chi_L(1-\frac{\chi_L}{\chi_G})$ by $\kappa(\hat{\theta})=\int_0^{\chi_G}\delta(\chi_L,\hat{\theta})W_L(\chi_L,\chi_G)d\chi_L$ when $\chi_L<\chi_G$ and zero otherwise, and where $\Omega_m$ is the current day matter density parameter. The comoving distance $\chi$ is given in units of $c/H_0$, where $H_0$ is the current day Hubble constant. The 3D matter power spectrum is then defined from the convergence as
\begin{equation}
\langle \tilde{\kappa}(\bm{\ell_1})\tilde{\kappa}(\bm{\ell_2})\rangle=(2\pi)^2\delta^{D}(\bm{\ell_1}+\bm{\ell_2})P(\ell_1),\label{eq:corr}
\end{equation}
where $\langle\cdots\rangle$ denotes the ensemble average and $\delta^{D}(\bm{\ell})$ is the Dirac delta function. Under the Limber approximation, we can relate the 2D angular cross-power spectrum between the $i$th and $j$th redshift bins to the 3D matter power spectrum as
\begin{equation}
C^{\alpha\beta}_{ij}(\ell)=\int_0^{\chi}\frac{W^{\alpha}_{i}(\chi')W^{\beta}_{j}(\chi')}{\chi'^2}P(\ell;\chi')d\chi',\label{eq:ps}
\end{equation}
where $\alpha,\beta\in G,\phi$, where G and $\phi$ represent galaxy and CMB lensing, respectively. For galaxy lensing, the weighting function $W^G_{i}$ is given by
where $\alpha,\beta\in G,\phi$, where G and $\phi$ represent galaxy and CMB lensing, respectively. For galaxy lensing, the weighting function $W^G_{i}$ is given by
\begin{equation}
W^G_{i}(\chi)=\int_0^{\chi}W_L(\chi',\chi)f_i(\chi')d\chi',\label{eq:weighting}
\end{equation}
where $f_i(\chi)$ is the comoving distribution of galaxies in the $i$-th redshift bin. For a CMB source, $f_i(\chi)\approx \delta^{D}(\chi-\chi^{*})$, where $\chi^{*}$ is the comoving distance to the surface of last scattering. This simplifies Eq. \ref{eq:weighting} to be
\begin{equation}
W^{\phi}(\chi)=\frac{3}{2}\Omega_m(1+z)\chi(1-\frac{\chi}{\chi^{*}}),\label{eq:weighting2}
\end{equation}
where we have suppressed the redshift bin denotation, since the source is at a single redshift.

\subsection{CMB lensing-galaxy lensing ($\phi I$) contaminant}
In cosmic shear studies using galaxy shapes, the impact of the $GI$ intrinsic alignment correlation on lensing information is well-known, and significant work has gone into mitigating it (e.g. \cite{hirata,17,6j,18b,9a,9b,19,20a,20b,20c,zhang,23,troxel,troxelc,heymans}). While there is no intrinsic alignment to impact CMB lensing auto-correlations, there should be a similar correlation between the lensing information encoded in the deflection and polarization of CMB photons and the intrinsic alignment of foreground galaxies, as well, through the CMB lensing-galaxy lensing cross-correlation. This becomes clear from the following physical argument.

{First, consider that some galaxy shape is composed of both an intrinsic shape component ($I$) and a component due to gravitational lensing ($G$). In the case of CMB lensing-galaxy cross-correlations, this galaxy can be at very high redshift. It is clear that the lensing information in $G$ should be correlated with the CMB lensing signal, since both the galaxy with shear $G$ and the CMB will be lensed by the same foreground structures. The intrinsic component ($I$) is instead influenced (or aligned) by the local tidal action of the surrounding matter structure. This matter structure will also contribute to the gravitational lensing of background photons from the CMB. Just as this intrinsic component of the shear $I$ is anti-correlated with the lensing of background galaxies, it should thus also be correlated to some degree with the lensing of the CMB. This can alternatively be expressed analytically, in analogy to the process for the $GI$ and $II$ terms in galaxy lensing, as a measured galaxy shape, $\gamma_{obs}=\gamma+\gamma^I$, being composed of both a lensing and intrinsic alignment contribution. The observed CMB lensing-galaxy lensing cross-correlation would then be
\begin{equation}
\langle\phi\gamma_{obs}\rangle=\langle\phi(\gamma+\gamma^I)\rangle=\langle\phi\gamma\rangle+\langle\phi\gamma^I\rangle.
\end{equation}
We will label these two resulting correlations $\phi G$ and $\phi I$, respectively, while the total correlation is the observed CMB lensing-galaxy lensing signal.}

\begin{figure}
\begin{center}
\includegraphics[width=\columnwidth]{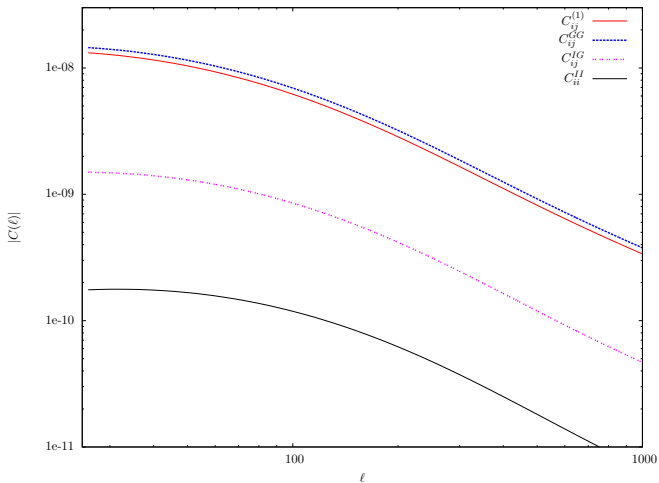}%
\caption{A comparison of the $GG$, $GI$, and $II$ components to the observed galaxy ellipticity-ellipticity spectrum $C^{(1)}$, as given in Eq. \ref{eq:observables} for a single-bin Stage IV weak lensing survey. The effect of $GI$ is a negative contamination of the galaxy lensing signal, which is about 10\% of $GG$ and significantly stronger than the $II$ contamination, which is very small for a deep galaxy lensing survey. This causes a total systematic decrease in the magnitude of the observed spectrum $C^{(1)}$ relative to the expected $GG$ signal. \label{fig:Phii}}
\end{center}
\end{figure}

\begin{figure}
\begin{center}
\includegraphics[width=\columnwidth]{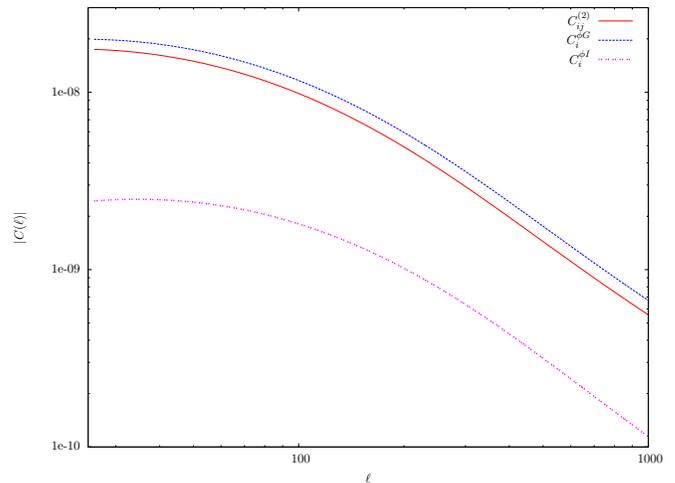}%
\caption{A comparison of the $\phi G$ and $\phi I$ components to the observed CMB lensing-galaxy lensing cross-spectrum $C^{(2)}$, as given in Eq. \ref{eq:observables} for a single-bin Stage IV weak lensing survey. The $\phi I$ component is negative and makes up about 15\% of the observed signal ($C^{(2)}$), creating a systematic decrease compared to the expected $\phi G$ lensing signal. This effect is stronger than the corresponding negative $GI$ contamination of the galaxy lensing signal, which is about 10\% of $GG$. The total magnitude of both $\phi G$ and $\phi I$ is larger than that of $GG$ and $GI$, and is consistent with the detection in \protect\cite{hand}.\label{fig:Phii2}}
\end{center}
\end{figure}

The cross-correlation of CMB lensing and galaxy lensing has been proposed as a method to constrain cosmology at intermediate redshifts, and thus we should consider the impact of such an intrinsic alignment contamination to the pure $\phi G$ lensing signal. In cosmic shear studies using galaxy shapes, the intrinsic alignment contamination $GI$ has been shown to contaminate the signal by up to $10\%$. We compare the resulting $GG$, $GI$, and $II$ signals for a single-bin Stage IV weak lensing survey in Figure \ref{fig:Phii}, with both $\phi G$ and $\phi I$ in Fig. \ref{fig:Phii2}. The various spectra are calculated as discussed in detail below. Both lensing signals have similar levels of contaminations due to the intrinsic alignment correlations on the order of 10-15\%. However, $\phi I$ (like $\phi G$) is slightly larger in magnitude than $GI$ ($GG$), and is fractionally about a 50\% stronger contaminant. We thus expect a similar or stronger level of bias in cosmological measurements due to the intrinsic alignment correlation in the CMB lensing-galaxy lensing cross-correlation as for the galaxy lensing auto-correlation, where it impacts constraints of the matter fluctuation amplitude, for example, at the 10\% level \cite{10} and the dark energy equation of state by up to 50\% \cite{9a,9b}. 
\noindent 
\noindent 
\section{Proposal to isolate and remove the intrinsic alignment $\phi I$ contamination}
In previous methods developed to self-calibrate the $GI$ correlation, complementary information in the form of the galaxy-ellipticity spectrum was used to isolate and remove the $GI$ contamination. Here we extend this process to calibrate the galaxy intrinsic alignment contamination to the CMB lensing-galaxy lensing cross-correlation. This presumes, of course, that one has successfully applied some method to measure or isolate the $GI$ galaxy ellipticity-IA cross-correlation, and requires overlapping measurements of the lensing of galaxy shapes and the CMB. A first detection of this CMB lensing-galaxy lensing cross-correlation has been made \citep{hand}, however, and we anticipate that much stronger detections will be possible with the design of overlapping fields in ongoing and future galaxy and CMB surveys.

To calibrate the $\phi I$ correlation, we proceed in a way that is analogous to the 2-point $GI$ self-calibration \citep{zhang}, where we first build a scaling relationship between the intrinsic alignment information in each observable. The two observable spectra of interest are the ellipticity-ellipticity correlation from galaxy surveys and the CMB lensing-galaxy lensing cross-correlation:
\begin{eqnarray}
C^{(1)}_{ij}(\ell)&=&C^{GG}_{ij}(\ell)+C^{IG}_{ij}(\ell)+C^{II}_{ij}\\\label{eq:observables}
C^{(2)}_{i}(\ell)&=&C^{\phi G}_{i}(\ell)+C^{I \phi}_{i}(\ell),
\end{eqnarray}
for $i<j$. We will ignore the $II$ correlation in what follows, as it is not a necessary part of calibrating the $\phi I$ cross-correlation. The $GI$ and $\phi I$ spectra can be expressed under the Limber approximation as
\begin{eqnarray}
C^{IG}_{ij}(\ell)&=&\int_0^{\chi}\frac{W^{G}_{j}(\chi')f_{i}(\chi')}{\chi'^2}P_{\delta I}(\ell;\chi')d\chi',\label{eq:psI}
\\
C^{I \phi}_{i}(\ell)&=&\int_0^{\chi}\frac{W^{\phi}(\chi')f_{i}(\chi')}{\chi'^2}P_{\delta I}(\ell;\chi')d\chi',
\end{eqnarray}
where $P_{\delta I}$ is the 3D matter-intrinsic alignment cross-spectrum. Assuming a sufficiently narrow comoving distribution of galaxies in each photo-z bin, these can be approximated as
\begin{eqnarray}
C^{IG}_{ij}(\ell)&\approx&W^G_{ij}\frac{P_{\delta I}(\ell;\chi_i)}{\chi_i^2},\label{eq:psIapprox}
\\
C^{I \phi}_{i}(\ell)&\approx&W^{\phi}_i\frac{P_{\delta I}(\ell;\chi_i)}{\chi_i^2},
\end{eqnarray}
where $W^G_{ij}=\int_0^{\chi}W^{G}_{j}(\chi')f_{i}(\chi')d\chi'$ and $W^{\phi}_i=\int_0^{\chi}W^{\phi}(\chi')f_{i}(\chi')d\chi'$. $W^{G}_{ij}$ is identical to the factor derived as part of the $GI$ self-calibration \citep{zhang}.
Combined, we can express $C^{I \phi}_{i}$ as a scaling of $C^{IG}_{ij}$
\begin{equation}
C^{I\phi}_i(\ell)\approx \frac{W^{\phi}_i}{W^G_{ij}}C^{IG}_{ij}(\ell).\label{eq:scaling1}
\end{equation}
Unlike the scaling relation in the $GI$ self-calibration, this scaling factor $W^{\phi}_i/W^G_{ij}$ can be simplified as
\begin{equation}
\frac{W^{\phi}_i}{W^G_{ij}}=\frac{\int_0^{\chi}(1+z_G)\chi_G(1-\frac{\chi_G}{\chi^{*}})f_{i}(\chi_G)d\chi_G}{\int_0^{\chi}\int_0^{\chi_G}(1+z_L)\chi_L(1-\frac{\chi_L}{\chi_G})f_i(\chi_L)f_{i}(\chi_G)d\chi_Gd\chi_L}.\label{eq:simpfrac}
\end{equation}
Though this process requires overlapping measurements in both CMB lensing and galaxy lensing surveys, we can see from this simplification that, unlike the $GI$ self-calibration, there is no need for priors on $\Omega_m$ and $H_0$ to evaluate the scaling relation, and it is instead dependent only on the redshift (or comoving) distribution of galaxies. It also doesn't require information about the galaxy bias, which reduces the impact of measurement errors in the calibration process compared to the $GI$ self-calibration.

\subsection{Accuracy of the scaling relationship between $\phi I$ and $GI$}
In order to probe the performance of such a calibration, we assume a Stage IV photometric weak lensing survey covering half the sky with a fully overlapping CMB lensing map. The redshift distribution for such a survey is given by 
\begin{equation}
f(z)=\frac{z^2}{2z_0^3}e^{-z/z_0},
\end{equation}
where $z_0=0.27$. The mean galaxy number density is 40 per arcminute$^2$, and the photo-z PDF is given by 
\begin{equation}
p(z|z^p)=\frac{1}{\sqrt{2\pi}\sigma_z}e^{-\frac{(z-z^p)^2}{2\sigma_z^2}},
\end{equation}
with $\sigma_z=0.05(1+z)$. In order to compute the non-linear matter power spectrum $P_{\delta}(k;\chi)$, we use the fitting formula of \cite{smith03}. The intrinsic alignment spectra, $P_{\delta I}(k;\chi)$ and $P_{I}(k;\chi)$, are calculated using the non-linear alignment model of \cite{9a}, which modifies the linear alignment model of \cite{hirata} on small scales by using the non-linear matter power spectrum. We estimate the spectrum amplitude $C_1$ through comparison to Fig. 2 of \cite{hirata}. The galaxies are split into redshift bins of width $\Delta z^p=0.2$, centered at redshifts $z^p=0.1+0.2 i$, where $i=1\dots 12$.

\begin{figure}
\begin{center}
\includegraphics[width=\columnwidth]{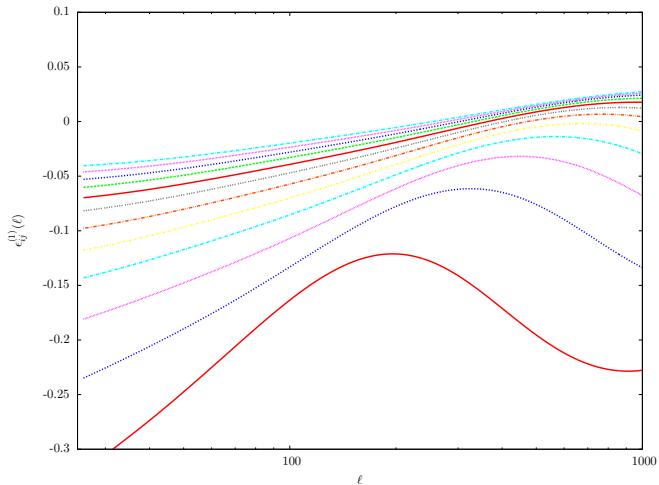}%
\caption{The inaccuracy of the scaling relationship in Eq. \ref{eq:scaling1} for each photo-z bin. The median photo-z ($z^p$) of the redshift bins in each curve increases going upward, such that $z^p=0.1+0.2i$ for $i=1$ (bottom curve) to $12$ (top curve). Compared to the $GI$ self-calibration of \protect\cite{zhang}, the inaccuracy is relatively high, ranging from 5-30\%. This corresponds to a potential reduction in the $\phi I$ contamination by a factor of 3-20, depending on photo-z bin.\label{fig:inaccuracy1}}
\end{center}
\end{figure}

There will be some systematic error induced in the CMB lensing-galaxy lensing cross-correlation measurement ($\phi$G) due to inaccuracies of the approximations used in Eqs. \ref{eq:scaling1}. We parameterize this inaccuracy as
\begin{equation}
\epsilon^{(1)}_{ij}=\left(\frac{W^{\phi}_i}{W^G_{ij}}\frac{C^{IG}_{ij}(\ell)}{C^{I\phi}_{i}(\ell)}\right)^{-1}-1.\label{eq:inaccuracy1}
\end{equation}
This inaccuracy is shown in Fig. \ref{fig:inaccuracy1}, for $i=j$. We find that the calibration of $\phi I$ from $GI$ performs with an inaccuracy of 5-30\% for these photo-z bins. This corresponds to a reduction of a factor of 3 in the lowest photo-z bin, though it is more typically a factor of 10 or more in higher redshift bins. The actual systematic error induced in the lensing measurement is then given by $\delta f_{ij}=\epsilon_{ij}f^I_{ij}$, where $f^I_{ij}=C^{I\phi}/C^{G\phi}$ is the fractional contamination of the lensing cross-spectrum. Typically, $f^I_{ij}\ll 1$. Thus, the systematic error is typically significantly less than $\epsilon_{ij}$. 
\noindent 
\subsection{Calibration of $\phi I$ from observables through the $GI$ self-calibration technique}
\begin{figure}
\begin{center}
\includegraphics[width=\columnwidth]{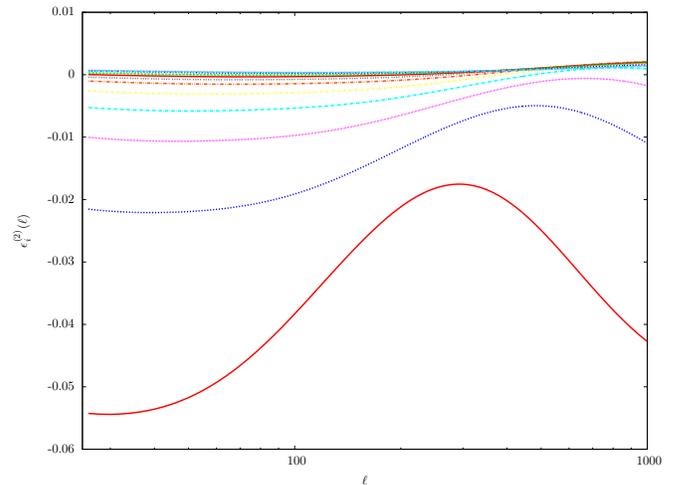}%
\caption{The inaccuracy of the scaling relationship in Eq. \ref{eq:scaling2} for each photo-z bin. The median photo-z ($z^p$) of the redshift bins in each curve increases going upward, such that $z^p=0.1+0.2i$ for $i=1$ (bottom curve) to $12$ (top curve). Compared to the $GI$ self-calibration of \protect\cite{zhang}, the $\phi I$ calibration from $C^{gI}$ performs much better, with inaccuracy ranging from effectively zero to about 5\%. This corresponds to a potential reduction in $\phi I$ down to detection limits for most photo-z bins.\label{fig:inaccuracy2}} 
\end{center}
\end{figure}

While the scaling relation in Eq. \ref{eq:scaling1} requires no information on the galaxy bias, it is possible to improve on the previous systematic estimates by considering the calibration of $\phi I$ as part of the $GI$ self-calibration process. In the $GI$ self-calibration, the galaxy overdensity-intrinsic ellipticity spectrum (gI) can be isolated from the galaxy overdensity-ellipticity spectrum, $C^{(3)}_{ii}=C^{gG}_{ii}+C^{gI}_{ii}$, using the estimator derived in \cite{zhang}. $C^{GI}_{ij}$ is then calculated from the scaling relation
\begin{equation}
C^{IG}_{ij}(\ell)\approx \frac{W^{G}_{ij}}{b_i \Pi_{ii}}C^{Ig}_{ii}(\ell),\label{eq:scaling2}
\end{equation}
where $b_i$ is the average galaxy bias in the $i$-th redshift bin and $\Pi_{ii}=\int^{\infty}_0 f^2_i(\chi)d\chi$. We can combine Eqs. \ref{eq:scaling1} \& \ref{eq:scaling2} into a single scaling relationship, which simplifies to
\begin{equation}
C^{I\phi}_i(\ell)\approx \frac{W^{\phi}_i}{b_i \Pi_{ii}}C^{Ig}_{ii}(\ell).\label{eq:scaling}
\end{equation}
We show the resulting inaccuracy of this relationship, 
\begin{equation}
\epsilon^{(2)}_{i}=\left(\frac{W^{\phi}_i}{b_i \Pi_{ii}}\frac{C^{Ig}_{ii}(\ell)}{C^{I\phi}_{i}(\ell)}\right)^{-1}-1,\label{eq:inaccuracy2}
\end{equation}
in Fig. \ref{fig:inaccuracy2}. This relationship is more accurate than Eq. \ref{eq:scaling1}, since the scaling information is directly mapped from the $i$-th photo-z bin to itself, and performs significantly better than the $GI$ self-calibration for the same reason. For low photo-z bins, we have only a 5\% inaccuracy, which reduces to effectively zero for high photo-z bins. This allows for an almost perfect mitigation of $\phi I$, with reductions down to detection limits in all photo-z bins.

\noindent 
\subsection{Residual statistical errors in the $\phi I$ calibration}
Any residual statistical errors in the estimate of $C^{\phi I}$ will propagate from the estimate of $C^{GI}$ as
\begin{equation}
\Delta C^{I\phi}_i(\ell)\approx \frac{W^{\phi}_i}{W^G_{ij}} \Delta C^{IG}_{ij}(\ell).\label{eq:errorscale}
\end{equation}
The relative fractional error will thus be unchanged between $GI$ and $\phi I$, given by the ratio of Eqs. \ref{eq:errorscale} \& \ref{eq:scaling1}.

If we use the self-calibration of $GI$ as an example, then
\begin{equation}
\Delta C^{I\phi}_i(\ell)\approx \frac{W^{\phi}_i}{b_i \Pi_{ii}} \Delta C^{Ig}_{ii}(\ell),\label{eq:errorscale2}
\end{equation}
where $\Delta C^{Ig}_{ii}(\ell)$ is given in \cite{zhang}. The residual statistical error is typically less than the minimum survey error for the $GI$ self-calibration, and thus it follows that the calibration of $\phi I$ is also safe from residual statistical errors due to the calibration process. Similarly, other conclusions regarding errors in the measurement of $b_i$, cosmological uncertainties, stochasticity, etc... hold for each, and we refer the reader to \cite{zhang} for more.

\section{Conclusion}
Measurements of CMB lensing and galaxy lensing are rapidly improving. Substantial overlap in coverage and increased statistical power in surveys which measure these quantities will allow us to explore cross-correlations between them, namely the CMB lensing-galaxy lensing signal, which provides us with an additional probe of structure at intermediate redshifts ($z\approx 1$). This cross-correlation has already been detected by \cite{hand}. One of the benefits to using CMB lensing is that the primary physical systematic which contaminates cosmic shear measurements from galaxy ellipticities, the intrinsic alignment of galaxies, is absent. However, we have shown here that this contamination is present in the CMB lensing-galaxy lensing cross-correlation at the level of $C^{\phi I}/C^{\phi G}\approx 15\%$, about $50\%$ stronger than the equivalent $GI$ contamination. We expect this to produce strong biases to any derived cosmological information, and must be accounted for in any precision cosmological study.

We also proposed a method to calibrate the $\phi I$ contamination by relating it to the $GI$ cross-correlation. This relationship is made more accurate when the $\phi I$ calibration is incorporated as part of the $GI$ self-calibration process that allows one to estimate $C^{\phi I}$ from $C^{gI}$, which can in principle be measured in a weak lensing survey. This calibration is more accurate than the estimation of $GI$, allowing for a nearly complete reduction of $\phi I$ by greater than a factor of $20$ for all photo-z bins. This process could totally alleviate the impact of bias due to the $\phi I$ contamination on cosmological information. {It also isolates a much deeper cross-correlation signal between the intrinsic alignment of galaxies and the CMB lensing signal. The CMB lensing signal contains information on structure in the universe back to the surface of last scattering, while this higher redshift intrinsic alignment signal contains information on the alignment of galaxies at much higher redshifts than can be probed in a typical galaxy lensing survey due to the higher redshift of the peak efficiency of the cross-correlation signal. This $\phi I$ signal thus will allow us to better study the evolution of the intrinsic alignment strength of galaxies within their dark matter halos over a much wider range of structure formation history.}

\acknowledgments
We thank Austin Peel and David Spergel for useful comments. We note that arXiv:1401.6018 had appeared on the archive a few days before our paper (arXiv:1401.7051) and dealing with the same cross-correlation term. There is partial overlap in some of the results, but the two works have been conducted independently and also present unique results. MI acknowledges that this material is based upon work supported in part by the National Science Foundation under grant AST-1109667. MT acknowledges that this work was supported in part by a NASA/TSGC graduate fellowship. 
\end{document}